\documentclass[onecolumn,floatfix,superscriptaddress,showpacs,showkeys,nofootinbib,preprint]{revtex4}
\usepackage{epsfig}
\usepackage{amssymb,latexsym,amsmath}
\newcommand{\eq}[1]{\begin{align} #1 \end{align}}

\begin{document}

\title{ Thermodynamical Consistency \\of  Excluded Volume Hadron Gas
Models}

\author{M.I. Gorenstein}

\affiliation{Bogolyubov Institute for Theoretical Physics, Kiev, Ukraine}
\affiliation{Frankfurt Institute for Advanced Studies, Frankfurt, Germany}

\begin{abstract}
The new excluded volume hadron gas model by Singh et al.
\cite{vdw1,vdw2,vdw2a,vdw3,vdw4,vdw5,vdw6} is critically
discussed. We demonstrate that in this model the results obtained
from relations between thermodynamical quantities disagree with
the corresponding results obtained by statistical ensemble
averaging. Thus, the model does not satisfy the requirements of
thermodynamical consistency.
\end{abstract}



\maketitle

\section{Introduction}
In order to take into account the effect of particle repulsion at
short distances the excluded volume procedure was proposed by van
der Waals in 1873. In a system with a fixed number of particles
$N$ this procedure corresponds to a substitution of the system
volume $V$ by the available volume $V- vN$, where $v$ is a proper
particle volume. This proposal was further supported by the
calculations within statistical mechanics   for a system of hard
spheres of radius $r$ at small particle number densities. In this
case, $v$ equals the particle hard-core volume $4\pi r^3/3$
multiplied by a factor of four. The excluded volume procedure
leads to the equation of state:
\eq{\label{p}
p\,(V~-~vN)~=~NT~,
}
or equivalently
\eq{\label{pnvdw}
p\,(1~-~vn)~=~nT~,
}
instead of the ideal gas relation
\eq{\label{pid}
p_{id}~=~n_{id}~T~.
}
In the above equations,  $p$ and $T$ are pressure and
temperature, respectively, and $n=N/V$ is particle number
density.

The hadron gas is a system with  conserved charges (baryonic
number, electric charge, strangeness), but with a variable number
of particles. Thus, for the hadron gas  the excluded volume
procedure (\ref{p}) with a variable number of particles is needed.
The first step  is to develop the excluded volume procedure in the
grand canonical ensemble (GCE),  as in the GCE a summation over
number of particles is performed.\footnote{ The excluded volume
procedure with a variable number of particles is also needed for
the hadron gas in the canonical and micro-canonical ensembles. For
these ensembles the conserved charges are exactly fixed, but the
number of particles still remains a variable quantity.} This step
appeared to be not trivial. The correct excluded volume procedure
in the system with a variable number of particles was first
introduced in Ref.~\cite{vdw}. Several other proposals
\cite{vd1,vd2,vd3} do not satisfy the thermodynamical relations.

Recently, the new excluded volume model has been developed by
Singh et al.~\cite{vdw1,vdw2,vdw2a,vdw3,vdw4,vdw5,vdw6} (we refer
to it as NEVM). In the grand canonical partition function the
authors substitute the system volume $V$ by the available volume
$V(1- \sum_j v_j\,n_j )$, where sum is taken over different
particle species, $n_j$ is the $j$-th particle number density, and
$v_j$ is the proper volume parameter of the $j$-th hadron.
Contrary to the statements in
Refs.~\cite{vdw1,vdw2,vdw2a,vdw3,vdw4,vdw5,vdw6} this procedure is
not thermodynamically consistent.
In order to prove this we consider the simplest example of the
system with a single particle type obeying Boltzmann statistics.

\section{Thermodynamical Consistency}
As a first step let us discuss general requirements of
thermodynamical consistency. In GCE the partition function is
given as the sum over states $i$:
\eq{\label{gcepart}
Z(V,T,\mu)~=~\sum_{i} \exp\left(\frac{\mu N_i-E_i}{T}\right)~,
}
where $N_i$ and $E_i$ are the number of particles and  system
energy in  $i$-state,  $T$ and $\mu$ are the system temperature
and chemical potential, respectively. The system pressure is
calculated as
\eq{\label{pTmu}
p~\equiv~\frac{T}{V}~\ln~Z(V,T,\mu)~,
}
whereas particle number density $n$ and energy density
$\varepsilon$ are found using the thermodynamical relations
\eq{\label{n}
n(T,\mu)~&=~\frac{\partial p}{\partial \mu}~,\\
\varepsilon(T,\mu)~&=~T\frac{\partial p}{\partial T} ~+~\mu
\frac{\partial p}{\partial \mu}~-~p~. \label{en}
%
}
Thus, thermodynamical average quantities for the number of
particles and system energy are equal to:
\eq{\label{N}
\overline{N}~&=~V\,n(T,\mu)~,\\
\overline{E}~&=~V\,\varepsilon(T,\mu)~.\label{E}
}
In order to fulfil the thermodynamical relations (\ref{n}) and
(\ref{en}) a correct structure of the partition function as a sum
over the system states $i$ is required.
The quantities  $T$ and $\mu$ may only enter  $Z(V,T,\mu)$
in the very definite way, i.e., only as given by
Eq.~(\ref{gcepart}).
Under these conditions the GCE statistical averages
\eq{
\langle N \rangle~&=~\frac{1}{Z(V,T,\mu)}~\sum_{i}~N_i~
\exp\left(\frac{\mu N_i-E_i}{T}\right)~, \label{N-cav}\\
\label{E-av}
\langle E \rangle~&=~\frac{1}{Z(V,T,\mu)}~\sum_{i}~E_i~
\exp\left(\frac{\mu N_i-E_i}{T}\right)~,
}
can be calculated using $T$- and $\mu$-derivatives as
\eq{
\langle N \rangle~&=
~\frac{1}{Z(V,T,\mu)}~\sum_{i}~T\frac{\partial }{\partial \mu}
\exp\left(\frac{\mu N_i-E_i}{T}\right)~
=~T\frac{\partial}{\partial \mu} \ln\,Z(V,T,\mu)~
=~V\frac{\partial p}{\partial \mu}~, \label{N-cav1}\\
\label{E-av1}
\langle E \rangle~&=~
\frac{1}{Z(V,T,\mu)}~\sum_{i}~\left(T^2\frac{\partial}{\partial T}~+~
T\mu\frac{\partial}{\partial \mu}\right)
\exp\left(\frac{\mu N_i-E_i}{T}\right)~\nonumber \\
&=~\left(T^2\frac{\partial}{\partial T}~+~
T\mu\frac{\partial}{\partial \mu}\right)\ln\,Z(V,T,\mu)
~=~V\left[T\frac{\partial p}{\partial T}~+
\mu\frac{\partial p}{\partial \mu}~-~p\right]~,
%
%
}
in agreement with thermodynamical relations (\ref{n}) and
(\ref{en}), i.e., the thermodynamical averages, $\overline{N}$
(\ref{N}) and $\overline{E}$ (\ref{E}), and statistical averages
$\langle N\rangle$ (\ref{N-cav}) and $\langle E\rangle$
(\ref{E-av}), are equal to each other:
\eq{\label{NE}
\langle N\rangle ~=~\overline{N}~,~~~~~~\langle E\rangle
~=~\overline{E}~.
}
\section{New Excluded Volume Model \cite{vdw1,vdw2,vdw2a,vdw3,vdw4,vdw5,vdw6}
}
NEVM \cite{vdw1,vdw2,vdw2a,vdw3,vdw4,vdw5,vdw6} assumes the
following expression for the GCE partition function
\eq{\label{part1}
Z(V,T,\mu)~&=~\sum_{N=0}^{\infty}\exp\left(\frac{\mu
N}{T}\right)~Z(V,T,N)~=~ \sum_{N=0}^{\infty}\exp\left(\frac{\mu
N}{T}\right)~\frac{(V~-~v\overline{N})^N}{N!}~z^N~
\nonumber \\
&=~ \exp\left[\exp(\mu/T)~\left(V~-~v\overline{N}\right)~z\right]~,
}
where
\eq{\label{z}
z(T)~=~\frac{g}{2\pi^2}\int_0^{\infty}k^2dk~
\exp\left[-~\frac{(k^2+m^2)^{1/2}}{T}\right]
}
is the so-called one-particle partition function. In
Eq.~(\ref{z}), $m$ and $g$ are the particle mass and the
degeneracy factor, respectively.
The quantity $\overline{N}$ in Eq.~(\ref{part1}) is an average
number of particles calculated with Eq.~(\ref{N}).
From
Eq.~(\ref{part1}) one finds the pressure:
\eq{\label{pres2}
p~\equiv~\frac{T}{V}~\ln~Z(V,T,\mu)~=
~T\,\exp\left(\mu/T\right)\,\left(1~-~vn\right)~z~
=~(1~-~vn)~p_{id}~.
}
Note that
\eq{\label{nid}
n_{id}(T,\mu)~=~ \exp(\mu/T)\,z(T)
}
is the particle number density in the ideal gas (i.e., at $v=0$),
and the ideal gas pressure $p_{id}$ is given by Eq.~(\ref{pid}).
The particle number density $n(T,\mu)$ is calculated in NEVM
\cite{vdw1,vdw2,vdw2a,vdw3,vdw4,vdw5,vdw6} using the
thermodynamical relation (\ref{n}), which together with
Eq.~(\ref{pres2}) yields:
\eq{\label{n2}
n~=~\left(1~-~vn~-~vT\frac{\partial n}{\partial
\mu}\right)~n_{id}~.
}

The pressure (\ref{pres2}) and particle number density (\ref{n2})
in NEVM are connected by the thermodynamical relation (\ref{n}).
Note, however, that in NEVM relation (\ref{pres2}) which connects
$p$ and $n$ is different from that given by Eq.~(\ref{pnvdw}).
The thermodynamical relations (\ref{n}) and (\ref{en}) do not
correspond in NEVM to correct statistical averaging of the number
of particles (\ref{N-cav}) and system energy (\ref{E-av}).
The canonical partition function in (\ref{part1}) reads:
\eq{\label{Zce}
Z(V,T,N)~=~\frac{(V~-~v\overline{N})^N~z^N}{N!}~,
}
with $\overline{N}$ given by Eq.~(\ref{N}).
A presence of $\overline{N}$, which is a function of $T$ and
$\mu$, in the partition function (\ref{part1}) destroys the
correct structure (\ref{gcepart}) of the GCE partition function,
thus, it leads to a violation of the thermodynamical consistency.
The thermodynamical relations (\ref{n}) and (\ref{en}) do not
correspond to correct statistical averaging in NEVM, i.e.,
\eq{\label{EE}
\langle N\rangle~\neq~\overline{N}~,~~~~~\langle
E\rangle~\neq~\overline{E}~.
}
Therefore, the thermodynamical relations (\ref{n}) and (\ref{en})
used in NEVM to calculate particle number density and energy
density are in contradiction with statistical averages
(\ref{N-cav}) and (\ref{E-av}). This is because a presence of
$\overline{N}=Vn(T,\mu)$ in the partition function (\ref{part1}),
which leads to the redundant terms in Eqs.~(\ref{N-cav1}) and
(\ref{E-av1}). These incorrect terms come from the derivatives of
$\overline{N}$ in respect to $\mu$ in (\ref{N-cav1}) and in
respect to  $T$ in (\ref{E-av1}).

\section{Excluded Volume Hadron Gas Model \cite{vdw}}
The excluded volume model \cite{vdw} is based on
the equation for the grand canonical partition function:
\eq{\label{evm}
Z(V,T,\mu) ~=~ \sum_{N=0}^{\infty}\exp\left(\frac{\mu
N}{T}\right)~\frac{(V~-~v N)^N}{N!}\,\theta(V-v N)~z^N~.
}
The form of the available volume $(V-vN)$  with the theta function
$\theta(V-vN)$ instead of $V-v\overline{N}$ in (\ref{part1})
causes some technical problems for the summation in (\ref{evm})
over $N$. On the other hand, just this form of the available
volume makes the formulation thermodynamically consistent. It is
easy to see that the thermodynamical relations (\ref{n}) and
(\ref{en}) are in agreement with statistical averaging
(\ref{N-cav}) and (\ref{E-av}) for the partition
function~(\ref{evm}). In calculations one uses the Laplace
transform of (\ref{evm}):
\eq{\label{LT}
\hat{Z}(s,T,\mu)&
=\int_{0}^{\infty}dV\exp(Vs)\,Z(V,T,\mu)=\sum_{N=0}^{\infty}
\frac{[z\,\exp(\mu/T)]^N}{N!}\int_{vN}^{\infty}
dV(V-vN)^N \exp(-sV) \nonumber \\
&
=~\frac{1}{s}\,\sum_{N=0}^{\infty}
\left[\frac{\exp(-v\,s)\,z\,\exp(\mu/T)}{s}\right]^N~=~
\Big[s~-~\exp(-~vs)~n_{id}(T,\mu)\Big]^{-1}~.
}
The system pressure in the thermodynamic limit can be then found
from the farthest-right singularity $s^*$ of $\hat{Z}$ (\ref{LT})
as a function of $s$:
\eq{\label{press1}
p(T,\mu)~=~\lim_{V\rightarrow \infty}\frac{T}{V}
~\ln Z(V,T,\mu)~=~T\,s^*(T,\mu)~.
}
The singularity $s^*$ is the pole singularity for (\ref{LT}),
which  leads to the transcendental equation for the pressure:
\eq{\label{press2}
p~=~\exp\left(-\, vp/T \right)~p_{id}~.
}
Using the thermodynamical relation (\ref{n}) one finds
\eq{\label{n-evm}
n~=~\frac{n_{id}~ \exp(-vp/T)}{1~+~v\,n_{id}\,\exp(-vp/T)}~.
}
Note that Eqs.~(\ref{press2}-\ref{n-evm}) are different from
Eqs.~(\ref{pres2}) and (\ref{n2}) of NEVM
\cite{vdw1,vdw2,vdw2a,vdw3,vdw4,vdw5,vdw6}. Combining
Eqs.~(\ref{press2}-\ref{n-evm}) and using the relation
$p_{id}=Tn_{id}$ one obtains after simple algebra
which coincides with Eq.~(\ref{pnvdw}).

\section{Closing remarks}

There are great difficulties in a formulation of statistical
theory of hadrons from the first principles. Models in this field
are of a phenomenological origin. An important constrain in model
formulation is the requirement of thermodynamical consistency. The
thermodynamical consistency can be violated in different ways. Two
examples are illustrated by considering the excluded volume model.
In the first example, the inconsistency appears when one
substitutes the system volume by the available volume in  all
ideal gas thermodynamical functions. This was done, e.g., in
Refs.~\cite{vd1,vd2,vd3}. The available volume is a function of
temperature and chemical potential, thus, the redundant terms
appear, and  the  thermodynamical relations (\ref{n},\ref{en}) are
not fulfilled after this substitution. The second example of the
inconsistency can be seen in the new excluded volume model of
Refs.~\cite{vdw1,vdw2,vdw2a,vdw3,vdw4,vdw5,vdw6}. In this model
one substitutes the system volume by the available volume in the
GCE partition function. Then the available volume is used to
calculate the pressure function only. Other thermodynamical
functions are calculated using the thermodynamical relations
(\ref{n},\ref{en}).
However, the available volume being a function of $T$ and $\mu$,
destroys the correct structure of the GCE partition function. This
leads to a violation of the thermodynamical consistency, i.e.,
results obtained from thermodynamical relations (\ref{n},\ref{en})
and from statistical averaging are different.

Several comments  concerning the excluded volume
model \cite{vdw} should be added. The (effective) proper volume
$v$  used in this model is  a model parameter. For a ``rigid''
balls it is a ball volume multiplied by a factor of 4. For
``soft'' hadrons this factor will have a value between 1 and 4.
However, in most cases the proper volume $v$ is treated as a free
parameter fitted to data.

A simplest case of a gas of identical particles was considered in
the above presentation for didactic reasons. The consistent
formulation of the excluded volume hadron gas model can be easily
extended to many particle species with different proper volume
parameters (see, for example, Ref.~\cite{GY}).

The effect of relativistic contraction is expected to be small in
the hadron gas. This is because the hadron gas temperature is
significantly smaller than almost all   hadron masses and thus
hadrons are non-relativistic.   This is not true for pions, their
masses are comparable to temperature. But a predominant majority
of pions comes from resonance decays, and  resonance masses are
again significantly larger  than the gas temperature. Most
important relativistic effect in the hadron gas model is a
variable number of hadrons. But this feature is exactly the main
point of the consistent formulation (\ref{evm}-\ref{n-evm}) of the
excluded volume hadron gas model.

\begin{acknowledgments}
I would like to thank M.~Ga\'zdzicki and J.~Rafelski for
discussions and comments.  This work was supported by Humboldt
Foundation and by the Program of Fundamental Research of the
Department of Physics and Astronomy of NAS, Ukraine.
\end{acknowledgments}

\end{document}